\documentclass[12pt]{article}
\usepackage{a4wide,amssymb,cite}
\pdfoutput=1

\usepackage{a4wide,amssymb,graphicx}
\usepackage{epsfig}
\usepackage[usenames,dvipsnames]{color}
\usepackage{slashed}
\usepackage{amssymb,cite,graphicx}
\usepackage{slashed}
\usepackage{amsmath,bm,bbm}
\usepackage{amsfonts}
\usepackage[titletoc,title]{appendix}
\usepackage[small]{caption}
\usepackage[margin=1in]{geometry}
\usepackage[multiple]{footmisc}
\usepackage{mathtools}
\usepackage{slashed}
\usepackage[nottoc]{tocbibind}
\usepackage{xcolor}

\newcommand{\be}{\begin{equation}}
\newcommand{\ee}{\end{equation}}
\newcommand{\bea}{\begin{eqnarray}}
\newcommand{\eea}{\end{eqnarray}}

\def\circa#1{\,\raise.3ex\hbox{$#1$\kern-.75em\lower1ex\hbox{$\sim$}}\,}

\begin{document}

\begin{titlepage}

\rightline{CERN-TH-2019-172}

\begin{centering}
\vspace{1cm}
{\Large {\bf The Selfish Higgs and Reheating}} \\

\vspace{1.5cm}

{\bf Hyun Min Lee}
\vspace{.5cm}

{\it Department of Physics, Chung-Ang University, Seoul 06974, Korea.} 
\\ \vspace{0.2cm}
{\it CERN, Theory department, 1211 Geneva 23, Switzerland. }

(Email: hminlee@cau.ac.kr, hyun.min.lee@cern.ch)

\end{centering}
\vspace{2cm}

\begin{abstract}
\noindent
We consider the cosmological relaxation of the Higgs mass and the cosmological constant due to the four-form fluxes in four dimensions.  We present a general class of models with a singlet scalar field containing four-form couplings where  the Higgs mass is relaxed to a right value and the Universe reheats to a sufficiently high reheating temperature after the last membrane nucleation. We also discuss some of interesting features in the cases of singlet scalar fields with non-minimal or minimal couplings to gravity and show how the new scalar fields can play a role for dark matter production.

\end{abstract}

\vspace{3cm}
%


\end{titlepage}

\section{Introduction}

The four-form flux provides an undetermined constant \cite{duff,witten,cc1,cc2}, enabling the cosmological constant to vary towards a small value.  The probability with the Euclidean action \cite{early} may prefer a small cosmological constant among the distribution of values with different flux parameters.
Moreover, the four-form flux can be changed in the process of creating membranes \cite{membrane}, with a tunneling probability between two configurations with cosmological constants differing by one unit \cite{tunneling}.

An interesting proposal was made recently for relaxing the cosmological constant and the Higgs mass parameter to observed values by the same four-form fluxes \cite{Giudice,Kaloper}.  A dimensionless coupling between the four-form flux and the Higgs field \cite{hierarchy,Higgscan} was introduced such that the flux parameter is scannable in steps of weak-scale value to relax the Higgs mass parameter to a correct value without a fine-tuning, whereas the anthropic argument is relied upon for obtaining the observed cosmological constant \cite{anthropic,Giudice,Kaloper}. The scanning of the Higgs mass parameter stops at a right value for electroweak symmetry breaking as the tunneling probability from the dS phase just after the last membrane nucleation and the AdS phase is exponentially suppressed. 

A non-minimal four-form coupling to gravity was introduced recently by the author as the minimal possibility for a successful reheating with the four-form flux \cite{hmlee}. Moreover, both the non-minimal four-form coupling to gravity and the four-form coupling to a pseudo-scalar inflaton \cite{inflation} 
were considered by the same author to show that a successful chaotic inflation with spontaneously broken shift symmetry is achieved \cite{hmleeinf}. 

In this article, we consider the general scenarios of the four-form flux with a singlet scalar degree of freedom for relaxing the Higgs mass parameter as well as the cosmological constant. 
Moreover, we discuss the reheating dynamics from the singlet scalar field which has the scalar potential with a four-form flux dependent minimum. We illustrate this in models containing the non-minimal four-form couplings to gravity introduced in \cite{hmlee}, or the four-form couplings to a pseudo-scalar field or a complex scalar field.
We show that the minimum of the scalar potential changes after the four-form flux relaxation and the natural initial condition for a successful reheating can be set after the last membrane nucleation. We discuss how the new scalar fields introduced with non-minimal or minimal couplings to gravity can also play a role of the mediator for dark matter production.

The paper is organized as follows.
We begin with the relaxation mechanism with the four-form flux in the SM minimally coupled to gravity and describe the effective theory for realizing the reheating process. Then, we present concrete examples for reheating with the non-minimal four-form coupling to gravity or the four-form couplings to singlet scalar fields and give the detailed discussion on reheating and dark matter production in each of the examples. Next, conclusions are drawn.

\section{The Relaxation mechanism with four-form flux}

We consider a three-index anti-symmetric tensor field $A_{\nu\rho\sigma}$ and its four-form field strength   $F_{\mu\nu\rho\sigma}=4\, \partial_{[\mu} A_{\nu\rho\sigma]}$.
Then, the most general Lagrangian with four-form field couplings in the SM are composed of various terms as follows,
\bea
{\cal L} = {\cal L}_0 +{\cal L}_{\rm int}+ {\cal L}_S +{\cal L}_L+ {\cal L}_{\rm memb} \label{full}
\eea
with
\bea
 {\cal L}_0 &=&  \sqrt{-g} \Big[\frac{1}{2}R  -\Lambda -\frac{1}{48} F_{\mu\nu\rho\sigma} F^{\mu\nu\rho\sigma} -  |D_\mu H|^2-V(H)\Big], \label{L0} \\
 {\cal L}_{\rm int} &=& \frac{c_2}{24} \,\epsilon^{\mu\nu\rho\sigma} F_{\mu\nu\rho\sigma} \, |H|^2,  \label{Lagint} \\
 {\cal L}_S &=&\frac{1}{6}\partial_\mu \bigg[\Big( \sqrt{-g}\,  F^{\mu\nu\rho\sigma} -c_2 \epsilon^{\mu\nu\rho\sigma}  |H|^2 \Big)A_{\nu\rho\sigma} \bigg],  \\
 {\cal L}_L &=& \frac{q}{24}\, \epsilon^{\mu\nu\rho\sigma} \Big( F_{\mu\nu\rho\sigma}- 4\, \partial_{[\mu} A_{\nu\rho\sigma]} \Big),  \label{LL} \\
 {\cal L}_{\rm memb}&=& \frac{e}{6} \int d^3\xi\,  \delta^4(x-x(\xi))\, A_{\nu\rho\sigma} \frac{\partial x^\nu}{\partial \xi^a} \frac{\partial x^\rho}{\partial \xi^b} \frac{\partial x^\sigma}{\partial \xi^c} \,\epsilon^{abc}. 
\eea
Here, the Higgs potential in the SM is given by
\bea
V(H) = -M^2 |H|^2 +\lambda_H |H|^4. 
\eea
We note that $c_2$ is a imensionless parameter for the four-form flux to the Higgs \cite{hierarchy,Giudice,Kaloper,hmlee},  taken to be positive in the later discussion without loss of generality. 
 ${\cal L}_S$ is the surface term necessary for the well-defined variation of the action with the anti-symmetric tensor field \cite{cc2}, and $q$ in ${\cal L}_L$ (in eq.~(\ref{LL})) is the Lagrange multiplier, and $ {\cal L}_{\rm memb}$ is the membrane action coupled to  $A_{\nu\rho\sigma}$ with membrane charge $e$ \footnote{The membrane tension can be also introduced by $-T\int d^3\xi\,  \delta^4(x-x(\xi)) \sqrt{-g^{(3)}}$ where $g^{(3)}$ is the determinant of the induced metric on the membrane. }.
Here, $\xi^a$ are the membrane coordinates, $x(\xi)$ are the embedding coordinates in spacetime and $\epsilon^{abc}$ is the volume form for the membrane.

Using  the equation of motion for $F_{\mu\nu\rho\sigma}$ as follows,
\bea
F^{\mu\nu\rho\sigma}=\frac{1}{\sqrt{-g}}\, \epsilon^{\mu\nu\rho\sigma} \Big( c_2 |H|^2+q\Big),
\eea
and integrate out $F_{\mu\nu\rho\sigma}$, we obtain the full Lagrangian (\ref{full}) as
\bea
{\cal L} &=&\sqrt{-g} \Big[\frac{1}{2}R-\Lambda_{\rm eff}-  |D_\mu H|^2 +M^2_{\rm eff} |H|^2 -\lambda_{H,{\rm eff}} |H|^4 \Big] \nonumber \\
&&+ \frac{1}{6}\epsilon^{\mu\nu\rho\sigma} \partial_\mu q A_{\nu\rho\sigma} +\frac{e}{6} \int d^3\xi \, \delta^4(x-x(\xi))\, A_{\nu\rho\sigma} \frac{\partial x^\nu}{\partial \xi^a} \frac{\partial x^\rho}{\partial \xi^b} \frac{\partial x^\sigma}{\partial \xi^c} \epsilon^{abc}. \label{Lagfull}
\eea 
\bea
M^2_{\rm eff}(q) &=& M^2 - c_2\, q, \\
  \Lambda_{\rm eff} (q) &=& \Lambda + \frac{1}{2}\, q^2, \\
  \lambda_{H,{\rm eff}}&=&\lambda_H+\frac{1}{2}c^2_2. 
  \eea
As a result, the equation of motion for $A_{\nu\rho\sigma}$ makes the four-form flux $q$  dynamical, according to
\bea
\epsilon^{\mu\nu\rho\sigma} \partial_\mu q= -e\int d^3\xi \, \delta^4(x-x(\xi))\, \frac{\partial x^\nu}{\partial \xi^a} \frac{\partial x^\rho}{\partial \xi^b} \frac{\partial x^\sigma}{\partial \xi^c} \epsilon^{abc}.
\eea
The flux parameter $q$ is quantized in units of $e$ as $q=e\,n$ with $n$ being integer. 
Whenever we nucleate a membrane, we can decrease the flux parameter by one unit  such that both the Higgs mass and the cosmological constant can be relaxed into observed values in the end. 

The membrane is located at the boundary between two consecutive dS space configurations that are defined by the flux parameters and differ by one unit. Then, it is argued the tunneling probability between those configurations is given \cite{tunneling} by
\bea
{\cal P}(n+1\rightarrow n) \approx {\rm exp} \left( -\frac{24\pi^2M^4_P}{\Lambda_{n+1}}\right) \label{tunnel}
\eea
when $\Lambda_{n+1}\ll T^2/M^2_P$ where $T$ is the membrane tension. 
Therefore, the probability of changing the flux parameter by one unit becomes large in the early stage of the nucleation, but it becomes extremely suppressed at the last stage, making the Universe entering in a metastable state with a small cosmological constant \cite{tunneling,membrane,Giudice,Kaloper}.
There has been a more elaborate discussion on the tunneling probability, including the effects of the finite membrane tension in Ref.~\cite{hmlee}.

In addition to the relaxation of the cosmological constant with four-form fluxes, the Higgs mass parameter is also scanned at the same time. 
For $q>q_c$ with $q_c\equiv M^2/c_2$, the Higgs mass parameter $M^2_{\rm eff}<0$, so electroweak symmetry is unbroken, whereas for $q<q_c$, we are in the broken phase.  For $c_2={\cal O}(1)$ and the membrane charge $e$ of electroweak scale,  we can explain the observed Higgs mass parameter once the flux change stops at $q=q_c-e$ by the previous argument for the tunneling probability \cite{Giudice,Kaloper}.
For $\Lambda<0$, we can cancel a large cosmological constant by the contribution from the same flux parameter until $\Lambda_{\rm eff}$ takes the observed value at $q=q_c-e$, but we need to reply on an anthropic argument for that with $e$ being of order weak scale \cite{anthropic}. 

As compared to the original proposal by Bousso and Polchinski \cite{membrane} where multiple four-form fluxes were introduced for the scanning of the cosmological constant with the precision of the observed value, we rely on a single four-form flux of electroweak scale to get both the Higgs mass and the cosmological constant at  right scales in our scenarios. The scanning of the cosmological constant can also depend on other contributions such as other four-form fluxes and the latent heats after phase transitions, but we regard them as being included in the brane cosmological constant $\Lambda$. Just before the last membrane nucleation, the flux parameter takes $q=q_c$ for which the effective cosmological constant is given by $\Lambda_{\rm eff}(q_c)=\Lambda+\frac{1}{2}q^2_c$. 
Then, just after the last membrane nucleation, the flux parameter decreases further to $q=q_c-e$, resulting in $\Lambda_{\rm eff}(q_c-e)=\Lambda+\frac{1}{2}(q_c-e)^2$. Then, the change in the effective cosmological constant after the last membrane nucleation is given by $\Delta \Lambda_{\rm eff}=\frac{1}{2}q^2_c-\frac{1}{2}(q_c-e)^2\simeq eq_c$ for $e\ll q_c$, which is much larger than the observed cosmological constant, unlike the case in Ref.~\cite{membrane}, thus makes the reheating more efficient \cite{hmlee}, as will be discussed in detail in the later sections. In order to get $\Lambda_{\rm eff}(q_c-e)$ to the observed value, we only have to tune the bare cosmological constant $\Lambda$ against the flux contribution at the time of the last membrane nucleation.
Once the necessary tuning for the cosmological constant is achieved after the last membrane nucleation, the probability for a further tunneling to the vacuum with a negative cosmological constant is highly suppressed according to eq.~(\ref{tunnel}), so we are eventually relaxed in a metastable universe with the correct Higgs mass and cosmological constant. 

We note that if there might be other vacua with a different bare cosmological constant, it would be impossible to tune the effective cosmological constant to the observed value for the same values of the flux parameter and the membrane charge, so those vacua would not be a livable universe, in the spirit of the anthropic argument \cite{anthropic}. If the flux parameter and the membrane charge are variable for a tunable cosmological constant in the other vacua, the Higgs mass would not come right as observed. 
Therefore, in this work, we pursue the interesting possibility that the scanning of the Higgs mass stops at a right scale only when the cosmological constant has its present value.

We also remark that there is a need of reheating at the end of the membrane nucleation. Otherwise the Universe would be empty after the continuous exponential expansion in dS phases.
We give a schematic description of the reheating dynamics  in the following general form of the effective potential containing a singlet scalar field or inflaton $\phi$,
\bea
V(H,\phi) = V_{\rm eff}(H) + (k_1\phi^n+q+k_2)^2+ V_{\rm int}(\phi,H)
\eea
where $V_{\rm eff}(H)=-M^2_{\rm eff} |H|^2 +\lambda_{H,{\rm eff}} |H|^4$, and $k_1,k_2$ are constant parameters and $n$ is the positive integer, and $V_{\rm int}(\phi,H)$  is the interaction potential between the SM Higgs and the inflaton. Then, the minimum of the inflaton potential changes after each membrane nucleation, so it is natural to realize the initial displacement of the inflaton field just before the last membrane nucleation and set the initial condition for reheating.
We will discuss some explicit examples for the inflaton potential in the next sections.

\section{Reheating with non-minimal four-form coupling}

In this section, we discuss the minimal possibility for the relaxation of the Higgs mass and the cosmological constant and reheating the universe as well as producing dark matter particles. 
To that purpose, we add the non-minimal four-form coupling to gravity as well as  $R^2$ term  \cite{hmlee}, as follows,
\bea
 {\cal L}_{\rm non-minimal} &=& -\frac{c_1}{24} \,\epsilon^{\mu\nu\rho\sigma} F_{\mu\nu\rho\sigma} \,R +\sqrt{-g}\, \bigg(\frac{1}{2}\zeta^2 R^2\bigg) \label{nonmin}
\eea
with the corresponding surface term,
\bea
\Delta {\cal L}_S =\frac{c_1}{6}\partial_\mu \bigg(\epsilon^{\mu\nu\rho\sigma} R A_{\nu\rho\sigma} \bigg). \label{surface2}
\eea
The most general Lagrangian in quadratic gravity also contains $R_{\mu\nu\rho\sigma}^2$ as well as a Gauss-Bonnet term. The latter term is physically irrelevant because it is a topological invariant, and we ignore $R_{\mu\nu\rho\sigma}^2$ for a consistent description of gravity without a ghost problem \cite{hmlee}.

\subsection{Relaxation of Higgs mass}

Considering a dual description of the $R^2$ term  in terms of a real scalar field $\chi$ \cite{hmlee},
the full Lagrangian ({\ref{full}}) with eqs.~(\ref{nonmin}) and (\ref{surface2}) becomes 
\bea
{\cal L} = \sqrt{-g} \bigg[ \frac{1}{2}\,\Omega(H,\chi,q) R  -  |D_\mu H|^2 +M^2_{\rm eff} |H|^2 -  \lambda_{\rm eff} |H|^4 -\Lambda_{\rm eff} -\frac{1}{2} \chi^2 \bigg]  \label{Lagfull3}
\eea
with
\bea
\Omega(H,\chi,q)=1 + c_1 \Big(c_2 |H|^2+q\Big)+\sqrt{\zeta^2-c^2_1}\, \chi
\eea
Furthermore, making the field redefinition by
\bea
\sigma= c_2 |H|^2+q+\frac{\sqrt{\zeta^2-c^2_1}}{c_1}\, \chi,
\eea
we get $\Omega=1 + c_1\sigma$ and  rewrite eq.~({\ref{Lagfull3}})  as
\bea
{\cal L}_{\rm I} = \sqrt{-g} \bigg[ \frac{1}{2}\,(1+c_1\sigma ) R  -  |D_\mu H|^2-V(H,\sigma,q)  \bigg] \label{Lagfinal}
\eea
with
\bea
V(H,\sigma,q) = -M^2_{\rm eff} |H|^2 +\lambda_{H,{\rm eff}}  |H|^4 +\Lambda_{\rm eff} +\frac{1}{2}\,\frac{c^2_1}{\zeta^2-c^2_1} \Big(\sigma-c_2 |H|^2-q \Big)^2.
\eea
We impose $\zeta^2>c^2_1$ for the potential for a new scalar field $\sigma$ to be bounded from below, without a need of a higher dimensional term to stabilize the potential. 

Making a Weyl scaling of the metric by $g_{\mu\nu}=g^E_{\mu\nu}/\Omega$, we get the Einstein frame Lagrangian as follows,
\bea
{\cal L}_E 
=\sqrt{-g_E} \bigg[ \frac{1}{2} R(g_E) -\frac{3}{4}\,c^2_1\,\Omega^{-2}\,(\partial_\mu\sigma)^2 - \frac{1}{\Omega}\, |D_\mu H|^2- \frac{V(H,\sigma,q)}{\Omega^2} \bigg]. 
\eea
Then, for arbitrary field values of $\sigma$,  the canonical sigma field ${\bar\sigma}$ in Einstein frame is redefined by
\bea
\sigma = \frac{1}{c_1} \Big(e^{\sqrt{\frac{2}{3}} {\bar\sigma}}-1 \Big),
\eea
and the Einstein frame Lagrangian becomes
\bea
{\cal L}_E 
=\sqrt{-g_E} \bigg[ \frac{1}{2} R(g_E) -\frac{1}{2}(\partial_\mu{\bar\sigma})^2 - e^{-\sqrt{\frac{2}{3}}{\bar\sigma}}\, |D_\mu H|^2- V_E(H,{\bar\sigma}) \bigg]
\eea
with
\bea
V_E(H,{\bar\sigma})&=& \Lambda_{\rm eff}\, e^{-2\sqrt{\frac{2}{3}}{\bar\sigma}}+\frac{3}{4} m^2_{\bar\sigma} \bigg(1-(1+c_1 q)e^{-\sqrt{\frac{2}{3}}{\bar\sigma}}-c_1 c_2\, e^{-\sqrt{\frac{2}{3}}{\bar\sigma}} |H|^2  \bigg)^2 \nonumber \\
&&+ e^{-2\sqrt{\frac{2}{3}}{\bar\sigma}} \Big( -M^2_{\rm eff}|H|^2+  \lambda_{H,{\rm eff}}|H|^4\Big)
\eea
and 
with
\bea
m_{\bar\sigma}= \sqrt{\frac{2}{3}}\, \frac{M_P}{\sqrt{\zeta^2-c^2_1}}. \label{inflatonmass}
\eea
Here, assuming that the SM Higgs is stabilized at $\langle H\rangle=v/\sqrt{2}$ in each dS phase, we can rewrite the above sigma field potential as
\bea
V_E(\sigma) = V_0(q) + \bigg[\frac{3}{4}m^2_{\bar\sigma}\Big(1+c_1\Big(q+\frac{1}{2}c_2 v^2\Big)\Big)^2+\Lambda_{\rm eff}  \bigg] \Big(e^{-\sqrt{\frac{2}{3}}{\bar\sigma}}- e^{-\sqrt{\frac{2}{3}}{\bar\sigma}_{\rm m}(q)}\Big)^2 \label{finpot}
\eea
where
\bea
e^{-\sqrt{\frac{2}{3}}{\bar\sigma}_{\rm m}(q)}&=& \frac{3m^2_{\bar\sigma}(1+c_1 (q+\frac{1}{2}c_2 v^2))}{3m^2_{\bar\sigma}(1+c_1 (q+\frac{1}{2}c_2 v^2))^2+4\Lambda_{\rm eff}}, \\
V_0(q)&=&\frac{3m^2_{\bar\sigma} \Lambda_{\rm eff}}{3m^2_{\bar\sigma}(1+c_1 (q+\frac{1}{2}c_2 v^2))^2 +4\Lambda_{\rm eff}}.
\eea

We also obtain the approximate form of the potential (\ref{finpot}): for $\Lambda_{\rm eff}\gg m^2_{\bar\sigma}$,
\bea
V_E\simeq  \frac{3}{4} m^2_{\bar\sigma} +\Lambda_{\rm eff} \Big( e^{-\sqrt{\frac{2}{3}}{\bar\sigma}}- \frac{3}{4} \frac{m^2_{\bar\sigma}}{\Lambda_{\rm eff}}\, (1+c_1 q)\Big)^2;
\eea
for $\Lambda_{\rm eff}\ll m^2_{\bar\sigma}$, 
\bea
V_E\simeq  \frac{\Lambda_{\rm eff}}{(1+c_1 q)^2} +  \frac{3}{4} m^2_{\bar\sigma} (1+c_1 q)^2 \Big( e^{-\sqrt{\frac{2}{3}}{\bar\sigma}}-\frac{1}{1+c_1 q} \Big)^2.
\eea
In both limits, away from the minimum, the sigma field dependent potential would become easily dominant, making the sigma field settling into the minimum very quickly.  But, there is a crucial difference between the two cases. In the first case with $\Lambda_{\rm eff}\gg m^2_{\bar\sigma}$, we can scan mostly the effective mass of the sigma field  with the flux parameter. In the  second case with $\Lambda_{\rm eff}\ll m^2_{\bar\sigma}$, the scanning of the cosmological constant with the flux parameter becomes more apparent. 
As we decrease $\Lambda_{\rm eff}$ for the decreasing $q$, it is natural to enter the regime with  $\Lambda_{\rm eff}\ll m^2_{\bar\sigma}$ and scan the cosmological constant while the sigma field mass is little dependent on the flux parameter.

\subsection{Reheating}

Now we discuss the reheating process in more detail. 
The possibility of reheating during the next-to-last dS phase was discussed \cite{hmlee}, but the last dS phase must be short-lived, resulting in the open universe. The dilution of the negative spatial curvature would required an extra inflation, so we could not maintain the sufficient reheating temperature in the end.
Therefore, we review the case where reheating takes place after the Higgs mass and the cosmological constant are relaxed to the observed values, that is, after the last membrane nucleation. 

Just before the last nucleation, we need $q=M^2/c_2\equiv q_c$ and $v=0$, for which
\bea
e^{-\sqrt{\frac{2}{3}}{\bar\sigma}_{\rm m}(q_c)} &\approx &\frac{1}{1+c_1 q_c} \Big(1+\frac{4eq_c}{3m^2_{\bar\sigma}(1+c_1 q_c)^2} \Big)^{-1},  \label{min1}\\
V_0 (q_c) &\approx & \frac{3m^2_{\bar\sigma} e q_c}{3m^2_{\bar\sigma}(1+c_1 q_c)^2 +4e q_c}
\eea
where we used $\Lambda_{\rm eff}(q_c-e)=\Lambda+\frac{1}{2}(q_c-e)^2\simeq 0$ in the end, and
\bea
\Lambda_{\rm eff}(q_c)= \Lambda + \frac{1}{2} q^2_c = e\Big(q_c-\frac{1}{2}e\Big)\approx e q_c.
\eea
After the last nucleation, we have $V_0\approx 0$ and
\bea
e^{-\sqrt{\frac{2}{3}}{\bar\sigma}_{\rm m}(q_c-e)}\approx \frac{1}{1+c_1 (q_c-e+\frac{1}{2}c_2 v^2)}\approx\frac{1}{1+c_1 q_c}. \label{min2}
\eea

Suppose that the sigma field settles into the minimum of the potential before the last nucleation.
Then, after the last nucleation, the minimum of the potential is shifted from eq.~(\ref{min1}) to eq.~(\ref{min2}). Taking the initial condition just before the last nucleation to be the minimum of the potential for $q=q_c$, i.e. ${\bar\sigma}_i={\bar\sigma}_{\rm m}(q_c)$, we can obtain the sigma field potential after the last nucleation as
\bea
V_E(\sigma)&\approx& \frac{3}{4}m^2_{\bar\sigma} \Big(1+\frac{4eq_c}{3m^2_{\bar\sigma}(1+c_1 q_c)^2} \Big)^{-2} \Big(e^{-\sqrt{\frac{2}{3}}({\bar\sigma}-{\bar\sigma}_i)}-1-\frac{4eq_c}{3m^2_{\bar\sigma}(1+c_1 q_c)^2} \Big)^2.
\eea
As a result, the sigma field starts to oscillate at ${\bar\sigma}={\bar\sigma}_i$ with the initial potential energy, given by
\bea
V_i\equiv V_E({\bar\sigma}_i) =\frac{12(e q_c)^2 m^2_{\bar\sigma}}{(3 m^2_{\bar\sigma} (1+ c_1q_c)^2+4 eq_c )^2} \label{maxpot}
\eea
where the latter approximation is made for $c_1 q_c\lesssim 1$.
Here, we find that: for $m^2_{\bar\sigma}\ll eq_c$,  $V_i\approx \frac{3}{4}  m^2_{\bar\sigma}$; for $  m^2_{\bar\sigma}\gg eq_c$, $V_i\approx  \frac{4}{3} (e q_c)^2/[m^2_{\bar\sigma}(1+c_1 q_c)^2]$.
On the other hand, for $m^2_{\bar\sigma}=\frac{2}{3}\sqrt{2} eq_c/(1+c_1 q_c)^2 $, the initial potential energy is maximized to $V_i\approx 0.25(e q_c)/(1+c_1 q_c)^2$.
Thus, the maximum initial potential can be obtained for the inflaton mass of order $1\,{\rm TeV}$  for $e\sim(1\,{\rm TeV})^2$ and $q_c\sim M^2_P$, but a heavier inflaton mass is favored for a sufficiently high reheating temperature.

Then, the general maximum temperature of the Universe after inflation is given by $T_{\rm max}=\left( \frac{90 V_i}{\pi^2 g_*} \right)^{1/4}$ with eq.~(\ref{maxpot}),  thus becoming
\bea
T_{\rm max}&\simeq& 2.5\times 10^{10}\,{\rm GeV} \left(\frac{100}{ g_*} \right)^{1/4}  \left(\frac{eq_c}{(1\,{\rm TeV}\cdot M_P)^2}\right)^{1/4}  \nonumber \\
&&\quad\times \bigg(\frac{m^2_{\bar\sigma} M^2_P}{eq_c}\bigg)^{1/4}\left(1+\frac{3}{4}\left(\frac{m^2_{\bar\sigma}M^2_P}{eq_c}\right)(1+c_1 q_c/M^2_P)^2 \right)^{-1/2} 
\eea 
where we have reintroduced the Planck scale for dimensionality. 
In particular, for $m^2_{\bar\sigma}\gg eq_c$ and $c_1 q_c/M^2_P\lesssim 1$, the maximum reheating temperature becomes
\bea
T_{\rm max}&\simeq& 1.5\times 10^{9}\,{\rm GeV} \left(\frac{100}{ g_*} \right)^{1/4}  \left(\frac{eq_c}{(1\,{\rm TeV}\cdot M_P)^2}\right)^{1/2}  \left(\frac{380\,{\rm TeV}}{m_{\bar\sigma}}\right)^{1/2}.
\eea

Since the inflaton coupling couples to the SM Higgs through the non-minimal coupling to the four-form flux, the perturbative decay rate of the inflaton into two Higgs bosons is given by
\bea
\Gamma({\bar\sigma}\rightarrow hh)= \frac{3c^2_1 c^2_2}{64\pi}  \frac{m^3_{\bar\sigma}}{M^2_P}.
\eea
Moreover, the inflaton can decay into a pair of the other SM particles through the linear inflaton coupling to the trace of the energy-momentum tensor in Einstein frame \cite{inflaton}. For instance, the partial decay rate of the inflaton decaying into a pair of SM fermions is given by $\Gamma({\bar\sigma}\rightarrow f{\bar f})=m^2_f m_{\bar\sigma}/(48\pi M^2_P)$ \cite{inflaton}, which is much smaller than the above decay rate into two Higgs bosons for $m_{\bar\sigma}\gg m_h$. So, we can approximate the total decay rate of the inflaton by the decay mode into two Higgs bosons. 

Then, the reheating temperature is determined by the inflaton decay to be
\bea
T_{\rm RH} =\left(\frac{90}{\pi^2 g_*} \right)^{1/4} ( \Gamma_{\bar\sigma} M_P)^{1/2}=  10\,{\rm MeV}\left(\frac{100}{ g_*(T_{\rm RH})} \right)^{1/4} \Big(\frac{c_1}{1} \Big) \Big(\frac{c_2}{1} \Big) \Big(\frac{m_{\bar\sigma}}{380{\rm TeV}}\Big)^{3/2}.
\eea
In this case, the reheating temperature is much smaller than the maximum temperature, due to the double suppressions with the Planck scale and the inflaton mass. But, we can obtain a sufficiently high reheating temperature for the successful BBN. 
We note that for $m_{\bar\sigma}\geq 1.6\times 10^8\,{\rm GeV}$, the reheating temperature becomes identical to the maximum reheating temperature,  that is, $T_{\rm RH}=T_{\rm max}$. 

In order for the slow-roll inflation to take place before the last membrane nucleation, we need the inflaton mass to be $m_{\bar\sigma}\ll H_I=8\times 10^{13}\,{\rm GeV}(r/0.1)^{1/2}$ where $r$ is the tensor to scalar ratio during inflation.  
Therefore, from the inflaton mass in eq.~(\ref{inflatonmass}), we need $\zeta\gtrsim 2.5\times 10^{4}\, (0.1/r)^{1/2}$.
As a consequence, for $2.5 \times 10^4 (0.1/r)^{1/2}\lesssim \zeta\lesssim 5.2\times 10^{12}$,  a slow-roll inflation and an instantaneous reheating is possible at the same time.

\subsection{Dark matter production}

Suppose that the Lagrangian for the SM fermion $\psi$ and a Dirac fermion dark matter $\chi$ in Jordan frame is given, as follows,
\bea
{\cal L}_\chi =\sqrt{-g}\bigg[{\bar\psi} i\gamma^\mu \Big(D_\mu+\frac{1}{2}\omega^{ab}_\mu \sigma_{ab}\Big) \psi - m_\psi{\bar \psi} \psi+  {\bar\chi} i\gamma^\mu \Big(D_\mu+\frac{1}{2}\omega^{ab}_\mu \sigma_{ab}\Big) \chi - m_\chi{\bar \chi} \chi \bigg]. 
\eea
Then, in Einstein frame with the metric, $g^E_{\mu\nu}=\Omega\, g_{\mu\nu}$, we obtain the DM fermion Lagrangian \cite{inflaton} as
\bea
{\cal L}_\chi &=&\sqrt{-g_E}\bigg[{\bar\psi}' i\gamma^\mu \Big(D_\mu+\frac{1}{2}\omega^{ab}_\mu \sigma_{ab}\Big) \psi' -  \Omega^{-1/2}\,m_\psi{\bar \psi'} \psi' \nonumber \\ 
&&\quad+{\bar \chi}' i\gamma^\mu \Big(D_\mu+\frac{1}{2}\omega^{ab}_\mu \sigma_{ab}\Big) \chi' - \Omega^{-1/2}\,m_\chi {\bar \chi'}\chi'     \bigg]
\eea
where the SM fermion and the DM fermion are rescaled to $\psi'=\Omega^{-3/4} \psi$ and $\chi'=\Omega^{-3/4} \chi$, respectively. 
As a result, the canonical sigma field $\bar\sigma$ has Planck-suppressed couplings to the SM fermion and dark matter through the trace of the energy-momentum tensor,
\bea
{\cal L}_{{\bar\sigma},\rm int} = \frac{1}{\sqrt{6} M_P}\, {\bar\sigma}\, (m_\psi {\bar\psi'} \psi'+ m_\chi {\bar\chi}' \chi').
\eea
Consequently, the partial decay rate of the inflaton into a pair of fermion dark matter is given by
$\Gamma({\bar\sigma}\rightarrow {\bar\chi}'\chi')= \frac{m^2_\chi m_{\bar\sigma}}{48 \pi M^2_P}$ \cite{inflaton}.
Furthermore, the inflaton becomes a natural candidate for the mediator between dark matter and the SM particles, but dark matter could never be in thermal equilibrium due to small couplings. Then, dark matter can be produced non-thermally from the decay of the inflaton into a dark matter pair, resulting in the DM abundance  \cite{inflatondecay} as
\bea
Y_{\chi+{\bar\chi}}= \frac{3}{4}\, {\rm BR}_{\bar\sigma}\,\cdot \frac{T_{\rm RH}}{m_{\bar\sigma}} 
\eea
where the branching ratio of the inflation decaying into a pair of fermion dark matter is given by ${\rm BR}_{\bar\sigma}\simeq \Gamma({\bar\sigma}\rightarrow {\chi}'\chi')/ \Gamma({\bar\sigma}\rightarrow hh)=\frac{4}{9c^2_1 c^2_2}\, \frac{m^2_\chi}{m^2_{\bar\sigma}}$.
Finally, we get the DM relic density as
\bea
\Omega_{\chi+{\bar\chi}} h^2 = 0.12 \,\left(\frac{100}{ g_*(T_{\rm RH})} \right)^{1/4} \bigg( \frac{m_\chi}{2\,{\rm TeV}}\bigg)^3 \bigg(\frac{380\,{\rm TeV}}{m_{\bar\sigma}} \bigg)^{3/2} \,\frac{1}{|c_1c_2|}.
\eea

We note that in the case of a real scalar dark matter  $S$ with mass $m_S$, the inflaton couples to the dark matter similarly by the trace of the corresponding energy-momentum tensor. In this case, the branching ratio of the inflation decaying into a pair of scalar dark matter is given by ${\rm BR}_{\bar\sigma}\simeq \frac{4}{9c_1^2 c^2_2}$ for $m_{\bar\sigma}\gg m_S$, so the relic density for scalar dark matter is given by
\bea
\Omega_S h^2 =  0.12 \,\left(\frac{100}{ g_*(T_{\rm RH})} \right)^{1/4} \bigg( \frac{m_S}{55\,{\rm MeV}}\bigg) \bigg(\frac{m_{\bar\sigma}}{380\,{\rm TeV}}\bigg)^{1/2} \,\frac{1}{|c_1c_2|}.
\eea
A similar result can be obtained for vector dark matter too.

\section{Pseudo-scalar field with minimal couplings}

We discuss the relaxation of the Higgs mass and the cosmological constant in the case where a singlet pseudo-scalar or complex scalar with four-form couplings plays a role for reheating and dark matter production. 
In this section, we begin with the case of a singlet pseudo-scalar.

We introduce a pseudo-scalar field $\phi$ with the four-form coupling as 
\bea
{\cal L}_{\rm pseudo-scalar} =  -\frac{1}{2} (\partial_\mu\phi)^2-\frac{1}{2} m^2_\phi\phi^2 +\frac{\mu}{24} \,\epsilon^{\mu\nu\rho\sigma} F_{\mu\nu\rho\sigma} \, \phi,  
\eea
with the corresponding surface term,
\bea
 \Delta {\cal L}_S =-\frac{\mu}{6}\partial_\mu \bigg(\epsilon^{\mu\nu\rho\sigma}  \phi A_{\nu\rho\sigma} \bigg).
 \eea
 We note that the shift symmetry for the pseudo-scalar field is respected by the four-form couplings, but it is explicitly broken by the mass term $m^2_\phi$.
Then, after using the equation of motion for $F^{\mu\nu\rho\sigma}$ with the four-form couplings to both pseudo-scalar and Higgs fields, we obtain the $A_{\nu\rho\sigma}-$independent part of the Lagrangian as
\bea
{\cal L}_{\rm II}&=& \sqrt{-g} \bigg[\frac{1}{2}R-\Lambda-  |D_\mu H|^2 +M^2 |H|^2 -\lambda_H |H|^4 \nonumber \\
&&-\frac{1}{2} (\partial_\mu\phi)^2-\frac{1}{2} m^2_\phi\phi^2-\frac{1}{2} (\mu \phi + c_2 |H|^2+q)^2  \bigg].
\eea 

In this model, for a general flux parameter $q$, the SM Higgs and the pseudo-scalar are expanded around the vacuum as $\langle H\rangle=(0,v_H(q)+h)^T/\sqrt{2}$ and $\langle\phi\rangle=v_\phi+\varphi$, with
\bea
v_H(q)&=& \sqrt{\frac{M^2- c_2(q+\mu v_\phi)}{\lambda_H+\frac{1}{2} c^2_2}},  \label{vev1} \\
v_\phi(q) &=& -\frac{\mu}{\mu^2+m^2_\phi} \cdot\Big( \frac{1}{2} c_2 v^2_H + q\Big). \label{vev2}
\eea
The minimum of the potential is stable as far as $m^2_\varphi m^2_h> c^2_2 \mu^2 v^2_H(q)$, where $m^2_\varphi=m^2_\phi+\mu^2$ and $m^2_h=2\lambda_{H,{\rm eff}} v^2_H(q)$. 
On the other hand, the mass eigenvalues and the mixing angle $\theta(q)$ are given by
\bea
m^2_{h_{1,2}}= \frac{1}{2} (m^2_\varphi+m^2_h) \mp \frac{1}{2} \sqrt{(m^2_\varphi-m^2_h)^2+4c^2_2 \mu^2 v^2_H(q)},  \label{masses}
\eea
and
\bea
\tan2\theta(q) = \frac{2c_2\mu v_H(q)}{m^2_\varphi-m^2_h}.  \label{mixing}
\eea
We note that in the absence of an explicit breaking of the shift symmetry, that is, $m^2_\phi=0$, there is no relaxation of a large Higgs mass, due to the fact that the minimization of the pseudo-scalar potential cancels the flux-induced Higgs mass. Thus, it is crucial to keep the explicit breaking mass term to be nonzero.

\subsection{Reheating with pseudo-scalar field }

We find that the critical value of the flux parameter for a vanishing effective Higgs mass parameter or $v_H=0$ is given by
\bea
q_c=\frac{1}{c_2}\, \Big(M^2- c_2\mu v_\phi(q_c)\Big). \label{qcrit}
\eea
Then, solving eq.~(\ref{qcrit}) with eq.~(\ref{vev2}) for $q_c$, we get
\bea
q_c&=&\frac{\mu^2+m^2_\phi}{m^2_\phi}\, \frac{M^2}{c_2}, \\
v_\phi(q_c)&=& -\frac{\mu}{m^2_\phi} \, \frac{M^2}{c_2}\equiv v_{\phi,c}, \label{vevcrit}
\eea
and the cosmological constant at $q=q_c$ is given by
\bea
V_c&=& \Lambda+\frac{1}{2} \Big(\mu v_\phi(q_c) + q_c\Big)^2+\frac{1}{2} m^2_\phi v^2_\phi \nonumber \\
 &=&\Lambda + \frac{1}{2} \frac{m^2_\phi}{\mu^2+m^2_\phi}\, q^2_c.
\eea

On the other hand, electroweak symmetry is broken at $q=q_c-e$, for which
\bea
v_H(q_c-e)&=& \sqrt{\frac{|m^2_H|}{\lambda_{H,{\rm eff}}}} \equiv v, \\
v_\phi(q_c-e) &=& v_{\phi,c}    -\frac{\mu}{\mu^2+m^2_\phi} \cdot\Big( \frac{1}{2} c_2 v^2 -e\Big)\equiv v_{\phi,0} \label{vevtrue}
\eea
with $|m^2_H|\equiv M^2-c_2(q_c-e+\mu v_\phi)$, and  the cosmological constant at $q=q_c-e$ is tuned to a tiny value as observed,
\bea
V_0&=& \Lambda -\frac{1}{4} \lambda_{\rm eff} v^4 +\frac{1}{2} \Big(\mu v_{\phi,0} + q_c-e\Big)^2+\frac{1}{2} m^2_\phi v^2_{\phi,0} \approx 0.
\eea

Consequently, we find that the weak scale depends on various parameters in the model, as follows,
\bea
v^2 = \frac{m^2_\phi}{\mu^2+m^2_\phi} \left(\frac{ c_2\,  e}{\lambda_{H,{\rm eff}}-  \frac{1}{2}  \frac{c^2_2\mu^2}{\mu^2+m^2_\phi}} \right).
\eea
In particular, as far as $m_\phi\sim \mu$, the weak scale can be obtained for the membrane charge $e$ of a similar scale, insensitive to the values of $m_\phi$ and $\mu$. But, for $m_\phi\ll \mu$, we can take a larger value of $e$. 
Moreover,  from eqs.~(\ref{vevcrit}) and (\ref{vevtrue}), after the last membrane nucleation, the pseudo-scalar VEV is shifted  by
\bea
\Delta v_\phi= v_{\phi,c}-v_{\phi,0}=  -\frac{\mu}{\mu^2+m^2_\phi} \cdot\Big( \frac{1}{2} c_2 v^2 -e\Big). 
\eea
As a result, we can make use of the flux-induced deviation of the pseudo-scalar field for reheating, as will be discussed below.  
 
Just after the last membrane nucleation, the full potential can be rewritten as
\bea
V(h,\phi)=\frac{1}{4}\lambda_{H,{\rm eff}}\Big( h^2-v^2\Big)^2+ \frac{1}{2}(\mu^2+m^2_\phi)\Big(\phi-v_{\phi,0}+\frac{c_2\mu}{\mu^2+m^2_\phi} (h^2-v^2)\Big)^2.
\eea
Then, setting the initial value of $\phi$ just before the last nucleation to $\phi_i=v_{\phi,c}$ and $\phi=\phi_i+\varphi$, the above potential just after the last nucleation becomes
\bea
V(h,\varphi)=\frac{1}{4}\lambda_{H,{\rm eff}}\Big( h^2-v^2\Big)^2+ \frac{1}{2}(\mu^2+m^2_\phi)\Big(\varphi-\Delta v_\phi+\frac{c_2\mu}{\mu^2+m^2_\phi} (h^2-v^2)\Big)^2.
\eea
Therefore, at the onset of the pseudo-scalar oscillation, with the SM Higgs frozen to $h=v$, the initial vacuum energy for reheating is given by
\bea
V_i &\equiv&  \frac{1}{2}(\mu^2+m^2_\phi) (\Delta v_\phi)^2 \nonumber \\
&=& \frac{1}{2} \frac{\mu^2}{\mu^2+m^2_\phi}\, \Big(e- \frac{1}{2} c_2 v^2\Big)^2.
\eea
So, the initial vacuum energy is about $V_i\sim e^2$ for $\mu\sim m_\phi$. 

Consequently, the maximum temperature of the Universe after inflation would be
\bea
T_{\rm max}=\left( \frac{90 V_i}{\pi^2 g_*} \right)^{1/4}\simeq 55\,{\rm GeV} \left(\frac{V^{1/4}_i}{100\,{\rm GeV}} \right)\left(\frac{100}{ g_*} \right)^{1/4} \label{Tmax}
\eea 
From the $\varphi$ coupling to the Higgs, ${\cal L}\supset - \frac{1}{2}c_2\mu \varphi h^2$, and $m_\varphi= \sqrt{m^2_\phi+\mu^2}$, the perturbative decay rate of the pseudo-scalar field into two Higgs bosons is given by
\bea
\Gamma_\varphi= \frac{c^2_2 \mu^2}{32\pi m_\varphi} \left(1-\frac{4m^2_h}{m^2_\varphi} \right)^{1/2}.
\eea
Then, for $c_2={\cal O}(1)$ and $\mu\sim m_\varphi\gtrsim 0.16 v$ for $\theta^2\lesssim 0.1$ to be consistent with the Higgs data, we get $\Gamma_\varphi\sim 0.1 m_\varphi\gtrsim 0.01 v$, for which $\Gamma_\varphi\gg H$ at $T_{\rm max}$, so the reheating is instantaneous. Therefore, the reheating temperature is given by $T_{\rm max}$ as in eq.~(\ref{Tmax}).

\subsection{Dark matter production}

Suppose that the pseudo-scalar field has an axion-like coupling to a fermion dark matter $\chi$ by
\bea
{\cal L}_{\phi,\rm int} =i\,\frac{\phi}{f}\, {\bar\chi} \gamma^5 \chi. 
\eea
Then, thanks to the flux-induced Higgs portal coupling for the pseudo-scalar field, ${\cal L}\supset - \frac{1}{2}c_2\mu \varphi h^2$, with a Higgs mixing, as discussed previously,  the pseudo-scalar field can communicate between dark matter and the SM, with the same four-form flux couplings.  In this case, the direct detection cross section for fermion dark matter is suppressed by the momentum transfer between dark matter and nucleon, due to the chiral operator $\gamma^5$ in the mediator coupling for dark matter \cite{axionmed}. This interesting behavior is due to the fact that the four-form couplings to both pseudo-scalar and Higgs fields exist, violating the $CP$ symmetry.

Since the maximum reheating temperature is limited by about $T_{\rm max}=55\,{\rm GeV}$ in this model, dark matter heavier than $55\,{\rm GeV}$ automatically become non-relativistic, even if it is thermalized just after reheating, so the freeze-out process would follow immediately for WIMP-like dark matter.

As dark matter can annihilate into a pair of the SM particles through the pseudo-scalar or Higgs boson, so  indirect detection experiments and Cosmic Microwave Background measurements \cite{indirect} can constrain dark matter with weak-scale masses.

\section{Complex scalar field with minimal couplings}

In this section, we consider another class of reheating scenarios with a complex scalar field after the four-form relaxation of the Higgs mass and discuss the role of the complex scalar field for the mediator for dark matter.

To this, we introduce a singlet complex scalar field $\Phi$ with a global or local $U(1)$ symmetry and the four-form coupling as 
\bea
{\cal L}_{\rm complex-scalar} =  -|\partial_\mu\Phi|^2-m^2_\Phi |\Phi|^2-\lambda_\Phi |\Phi|^4 +\frac{\alpha}{24} \,\epsilon^{\mu\nu\rho\sigma} F_{\mu\nu\rho\sigma} \, |\Phi|^2,  
\eea
with the corresponding surface term,
\bea
 \Delta {\cal L}_S =-\frac{\alpha}{6}\partial_\mu \bigg(\epsilon^{\mu\nu\rho\sigma}  |\Phi|^2 A_{\nu\rho\sigma} \bigg).
 \eea
Then, after using the equation of motion for $F^{\mu\nu\rho\sigma}$ with the four-form couplings to both complex scalar and Higgs fields, we obtain the $A_{\nu\rho\sigma}-$independent part of the Lagrangian as
\bea
{\cal L}_{\rm III}&=& \sqrt{-g} \bigg[\frac{1}{2}R-\Lambda-  |D_\mu H|^2 +M^2 |H|^2 -\lambda_H |H|^4 \nonumber \\
&&  -|\partial_\mu\Phi|^2-m^2_\Phi |\Phi|^2-\lambda_\Phi |\Phi|^4-\frac{1}{2} (\alpha |\Phi|^2 + c_2 |H|^2+q)^2  \bigg].
\eea 

For a general flux parameter $q$, taking $\alpha>0$ and  $m^2_\Phi<0$, the singlet complex scalar field gets a VEV with $\langle\Phi\rangle=\frac{1}{\sqrt{2}} \, v_\phi$,
\bea
v_\phi(q)=\sqrt{\frac{-m^2_\Phi-\alpha q -\frac{1}{2}\alpha c_2 v^2_H(q)}{\lambda_{\Phi,{\rm eff}}}}, \label{phivev}
\eea
with  $\lambda_{\Phi,{\rm eff}}\equiv \lambda_\Phi+\frac{1}{2} \alpha^2$,
and the Higgs VEV is given by
\bea
v_H(q) = \sqrt{\frac{M^2-c_2 q- \frac{1}{2}\alpha c_2 v^2_\phi(q)}{\lambda_{H,{\rm eff}}}}.
\eea
The stability of the minimum is ensured for $4\lambda_{\phi,{\rm eff}} \lambda_{H,{\rm eff}}> (\alpha c_2)^2$.  The mass eigenvalues and the mixing angle are given by
\bea
m^2_{h_{1,2}} = \lambda_{\Phi,{\rm eff}} v^2_\phi(q) +  \lambda_{H,{\rm eff}} v^2_H(q) \mp \sqrt{( \lambda_{\Phi,{\rm eff}} v^2_\phi(q) -  \lambda_{H,{\rm eff}} v^2_H(q))^2+\alpha^2 c^2_2 v^2_\phi(q) v^2_H(q)},
\eea
and
\bea
\tan2\theta(q) = \frac{\alpha c_2 v_\phi(q) v_H(q)}{\lambda_{\Phi,{\rm eff}} v^2_\phi(q) -  \lambda_{H,{\rm eff}} v^2_H(q)}.
\eea

\subsection{Reheating with complex scalar field}

The critical value of the flux parameter for $v_H=0$ is given by
\bea
q_c= \frac{1}{c_2} \bigg(M^2 - \frac{1}{2} \alpha c_2 v^2_\phi \bigg). \label{crit2}
\eea
Then, solving eq.~(\ref{crit2}) with eq.~(\ref{phivev}) for $q_c$, we obtain
\bea
q_c&=&\frac{1}{\lambda_\phi} \bigg(\frac{\lambda_{\Phi,{\rm eff}}}{c_2}\, M^2 + \frac{\alpha}{2}\, m^2_\Phi \bigg), \\
v^2_\phi(q_c) &=& -\frac{1}{\lambda_\phi} \, \bigg(m^2_\Phi + \frac{\alpha}{c_2}\, M^2 \bigg)\equiv v^2_{\phi,c}.
\eea
So, for $v^2_\phi>0$, we need $m^2_\Phi<- \frac{\alpha}{c_2}\, M^2$ for $\lambda_\Phi>0$, then $q_c< \lambda_\Phi M^2/c_2$ ; $m^2_\Phi>- \frac{\alpha}{c_2}\, M^2$ for $\lambda_\Phi<0$, then $q_c> -|\lambda_\Phi| M^2/c_2$. For either $\lambda_\Phi>0$ or $\lambda_\Phi<0$, the magnitude of the flux parameter is bounded from above and the vacuum stability is ensured as far as $\lambda_{\Phi,{\rm eff}} =\lambda_\Phi+\frac{1}{2} \alpha^2>0$.
For $q=q_c$, the cosmological constant is given by
\bea
V_c &=& \Lambda+ \frac{1}{2} q^2_c -\frac{1}{4} \lambda_{\Phi,{\rm eff}} v^4_{\phi,c} \nonumber \\
&=& \Lambda+ \frac{1}{2} q^2_c -\frac{1}{4\lambda_{\Phi,{\rm eff}}} \, (m^2_\Phi+\alpha q_c)^2.
\eea

On the other hand, when electroweak symmetry is broken at the last membrane nucleation to $q=q_c-e$, the VEVs now become
\bea
v_H(q_c-e) &=&  \sqrt{\frac{|m^2_H|}{\lambda_{H,{\rm eff}}}} \equiv v, \\
v^2_\phi(q_c-e) &=&  v^2_{\phi,c} + \frac{\alpha}{\lambda_{\Phi,{\rm eff}} }\, \Big(e -\frac{1}{2} c_2 v^2 \Big) \equiv v^2_{\phi,0}
\eea
where $|m^2_H|\equiv M^2 -c_2 (q_c-e) -\frac{1}{2} \alpha c_2 v^2_{\phi,0}$. Then,  we can determine the electroweak scale in terms of various dimensionless couplings and the membrane charge as
\bea
v^2= \frac{\lambda_\Phi c_2 e}{\lambda_{\Phi,{\rm eff}} \lambda_{H,{\rm eff}}-\frac{1}{4}(\alpha c_2)^2 }.
\eea
Therefore, the electroweak scale is of order the membrane charge unless there is a tuning in the dimensionless parameters. 
As in the previous section, after the last membrane nucleation the singlet scalar VEV is shifted  by 
\bea
 v^2_{\phi,c}- v^2_{\phi,0} = -\frac{\alpha}{\lambda_{\Phi,{\rm eff}} }\, \Big(e -\frac{1}{2} c_2 v^2 \Big).
\eea

Just after the last membrane nucleation, the full potential for $\Phi= \frac{1}{\sqrt{2}}\, \phi$ and the SM Higgs  can be rewritten as
\bea
V(h,\phi)= 
\frac{1}{4}\lambda_{H,{\rm eff}}\Big( h^2-v^2\Big)^2+ \frac{1}{4}\lambda_{\Phi,{\rm eff}}\Big(\phi^2-v^2_{\phi,0}+\frac{\alpha c_2}{\lambda_{\Phi,{\rm eff}}} (h^2-v^2)\Big)^2.
\eea
Then, setting the initial value of $\phi$ just before the last nucleation to $\phi_i=v_{\phi,c}$ and $\phi=\phi_i+\varphi$, the above potential just after the last nucleation becomes
\bea
V(h,\varphi)=\frac{1}{4}\lambda_{H,{\rm eff}}\Big( h^2-v^2\Big)^2+ \frac{1}{4}\lambda_{\Phi,{\rm eff}}\Big(\varphi^2+2v_{\phi,c}\varphi+v^2_{\phi,c}-v^2_{\phi,0}+\frac{\alpha c_2}{\lambda_{\Phi,{\rm eff}}} (h^2-v^2)\Big)^2.
\eea
Therefore, at the onset of the singlet scalar oscillation, with the SM Higgs frozen to $h=v$, the initial vacuum energy for reheating is given by
\bea
V_i &\equiv&  \frac{1}{4}\lambda_{\Phi,{\rm eff}}  (v^2_{\phi,c}-v^2_{\phi,0})^2 \nonumber \\
&=& \frac{1}{2} \frac{\alpha^2}{\lambda_{\Phi,{\rm eff}} }\, \Big(e- \frac{1}{2} c_2 v^2\Big)^2.
\eea
So, the initial vacuum energy is about $V_i\sim e^2$ as in the previous section. 

Consequently, the maximum temperature of the Universe after inflation is similarly given by eq.~(\ref{Tmax}).
From the $\varphi$ coupling to the Higgs, ${\cal L}\supset -\alpha c_2 v_{\phi,c} \varphi h^2$, 
the perturbative decay rate of the singlet scalar into two Higgs bosons is given by
\bea
\Gamma_\varphi= \frac{\alpha^2 c^2_2 v^2_{\phi,c}}{8\pi m_\varphi} \left(1-\frac{4m^2_h}{m^2_\varphi} \right)^{1/2}.
\eea
Then, for $\alpha, c_2={\cal O}(1)$ and $m_\varphi\simeq \sqrt{2\lambda_{\Phi,{\rm eff}}} \, v_{\phi,0} = m_\varphi\gg m_h$ to be consistent with the Higgs data, we get $\Gamma_\varphi\sim 0.1 m_\varphi\gtrsim 0.01 v$, for which $\Gamma_\varphi\gg H$ at $T_{\rm max}$, so the reheating is instantaneous. Therefore, the reheating temperature is given by $T_{\rm max}$ as in eq.~(\ref{Tmax}).

\subsection{Dark matter production}

In the reheating scenarios with a complex scalar field, we assumed that there is a global or local $U(1)$ in the hidden sector, under which a fermion or scalar dark matter can be charged and become stable. 
If dark matter is a chiral fermion, it can get mass due to the spontaneous breaking of the $U(1)$ symmetry.

In this case, the dark Higgs from the complex scalar field and the dark gauge boson associated with the local $U(1)$ can play a role of the mediator for dark matter. In particular, in our model, there is a model-independent Higgs portal coupling induced by the four-form flux, which is ${\cal L}\supset -\alpha c_2 v_{\phi,c} \varphi h^2$, as noted in the above.  Then, we can correlate between the four-form couplings and the dark matter interactions.

Since the four-form couplings to both complex scalar and Higgs fields respect the same parity, there is no CP violation, which means that the dark Higgs has usual Higgs-portal interactions. As the reheating temperature is limited by about $T_{\rm max}=55\,{\rm GeV}$, a similar conclusion for the freeze-out process of dark matter applies as in the previous case with a pseudo-scalar mediator.
But, as the DM-nucleon elastic scattering is mediated by the Higgs-like scalars, the WIMP possibility is limited due to the strong bounds from direct detection experiments, apart from the resonances, which is at $m_\varphi= 2m_{\rm DM}$ \cite{Higgsportal}.

\section{Conclusions}

We provided new scenarios for solving the hierarchy problem in the SM where the four-form flux can relax not only the Higgs mass but also the cosmological constant to observable values and the reheating mechanism is naturally implemented by the extra four-form couplings.  We showed that the non-minimal four-form coupling to gravity or the four-form couplings to singlet scalar fields gives rise to a successful reheating of the Universe at the end of relaxation and we described how the new scalar fields can play a role of the mediator for dark matter production.

\section*{Acknowledgments}

The author would like to thank the Galileo Galilei Institute for Theoretical Physics for hospitality during which part of this work was done. 
The work is supported in part by Basic Science Research Program through the National Research Foundation of Korea (NRF) funded by the Ministry of Education, Science and Technology (NRF-2019R1A2C2003738 and NRF-2018R1A4A1025334).




\end{document}